\documentclass[conference,10pt]{IEEEtran}
\IEEEoverridecommandlockouts
\usepackage{cite}
\usepackage{amsmath,amssymb,amsfonts}
\usepackage{algorithmic}
\usepackage{graphicx}
\usepackage{textcomp}
\usepackage{xcolor}
\usepackage{balance}

\usepackage{multirow}
\usepackage{listings}
\usepackage{booktabs}
\usepackage[flushleft]{threeparttable}
\usepackage{hyperref}

\newcommand{\ra}[1]{\renewcommand{\arraystretch}{#1}}

\newcommand{\todo}[1]{{\it\color{red} TODO: #1}}
\newcommand{\tzjtodo}[1]{{\it\color{orange} TODO(Tzj): #1}}
\newcommand{\note}[1]{{\it\color{blue} #1}}
\newcommand{\ignore}[1]{#1}
\newcommand{\etal}{et al.}
\newcommand{\diff}{\textit{diff}}
\newcommand{\toolname}{CCRep}
\newcommand{\varianttoken}{CCRep$^\text{token}$}
\newcommand{\variantline}{CCRep$^\text{line}$}
\newcommand{\varianthybrid}{CCRep$^\text{hybrid}$}
\newcommand{\finaltodo}[1]{{\color{red} TODO: #1}}

\renewcommand{\todo}[1]{}
\renewcommand{\tzjtodo}[1]{}
\renewcommand{\note}[1]{}
\renewcommand{\finaltodo}[1]{}
\renewcommand{\ignore}[1]{}

\definecolor{diffstart}{RGB}{173, 175, 177}
\definecolor{diffincl}{RGB}{0, 110, 0}
\definecolor{diffrem}{RGB}{255, 50, 0}
\definecolor{orange}{RGB}{255, 165, 0}
\definecolor{javapurple}{rgb}{0.5,0,0.35} 

\def\BibTeX{{\rm B\kern-.05em{\sc i\kern-.025em b}\kern-.08em
    T\kern-.1667em\lower.7ex\hbox{E}\kern-.125emX}}
\begin{document}

\title{CCRep: Learning Code Change Representations via Pre-Trained Code Model and Query Back

}

\author{
\IEEEauthorblockN{Zhongxin Liu}
\IEEEauthorblockA{\textit{Zhejiang University} \\
Hangzhou, China \\
liu\_zx@zju.edu.cn}
\and
\IEEEauthorblockN{Zhijie Tang}
\IEEEauthorblockA{\textit{Zhejiang Univeristy} \\
Hangzhou, China \\
tangzhijie@zju.edu.cn}
\and
\IEEEauthorblockN{Xin Xia\IEEEauthorrefmark{1}\thanks{\IEEEauthorrefmark{1}Corresponding author.}}
\IEEEauthorblockA{\textit{Huawei} \\
China \\
xin.xia@acm.org}
\and
\IEEEauthorblockN{Xiaohu Yang}
\IEEEauthorblockA{\textit{Zhejiang University} \\
Hangzhou, China \\
yangxh@zju.edu.cn}
}

\maketitle

\begin{abstract}
Representing code changes as numeric feature vectors, i.e., code change representations, is usually an essential step to automate many software engineering tasks related to code changes, e.g., commit message generation and just-in-time defect prediction.
Intuitively, the quality of code change representations is crucial for the effectiveness of automated approaches.
Prior work on code changes usually designs and evaluates code change representation approaches for a specific task, and little work has investigated code change encoders that can be used and jointly trained on various tasks.
To fill this gap, this work proposes a novel \textbf{C}ode \textbf{C}hange \textbf{Rep}resentation learning approach named \textbf{\toolname{}}, which can learn to encode code changes as feature vectors for diverse downstream tasks.
Specifically, \toolname{} regards a code change as the combination of its before-change and after-change code, leverages a pre-trained code model to obtain high-quality contextual embeddings of code, and uses a novel mechanism named query back to extract and encode the changed code fragments and make them explicitly interact with the whole code change.
To evaluate \toolname{} and demonstrate its applicability to diverse code-change-related tasks, we apply it to three tasks: commit message generation, patch correctness assessment, and just-in-time defect prediction.
Experimental results show that \toolname{} outperforms the state-of-the-art techniques on each task.
\end{abstract} 

\section{Introduction}\label{sec:intro}

During software development, developers constantly perform code changes to implement new features, fix bugs, and maintain existing code (e.g., refactoring)~\cite{brudaru2008what}.
A code repository can be regarded as a sequence of ordered code changes.
A code change can be represented as the combination of the code versions before and after the change or as a flat text, such as \diff.

Analyzing and understanding code changes are important for a bulk of software engineering tasks.
For example, commit message generation~\cite{jiang2017auto} requires to summarize the content and intent of code changes, and vulnerability fix identification~\cite{zhou2021finding} needs to analyze vulnerability-related information in a code change.
To automate these tasks, a popular and effective paradigm is first encoding code changes into feature vectors, which are expected to capture the information related to the target task, and then leveraging machine learning or information retrieval (IR) techniques for automation~\cite{kim2008classifying,kamei2016studying,jiang2017auto,tian2020evaluating,he2021automated}.
Such feature vectors are referred to as code change representations.
Intuitively, the more precise the code change representations are, the less challenging the downstream learning or retrieval process will be.
In contrast, if the representations miss some key information, it would be very hard, if not impossible, to obtain good results.
Therefore, for code-change-related tasks, the quality of code change representations is critical to the effectiveness of automated approaches~\cite{hoang2020cc2vec}.

Many researchers have investigated code change representation techniques for specific downstream tasks.
Some studies converted code changes into numerical vectors based on manually crafted features, such as the sizes of code changes and the syntactic structures being changed~\cite{kim2008classifying,tian2012identifying,kamei2016studying,zeng2021deep}.
Another line of work leverages neural networks to learn code change representations on downstream tasks in an end-to-end manner~\cite{jiang2017auto,hoang2019deepjit,lin2021traceability,zhou2021finding}, i.e., learning-based techniques.
Compared to the techniques based on manually crafted features, learning-based approaches automatically learn representations from data and have shown to be more effective in many tasks~\cite{wang2021context,zhou2021finding,lin2021context,zhou2022spi}.
However, many of them adopt task-specific architectures and are trained from scratch, which makes it non-trivial to adapt them to other tasks, especially the tasks with only small datasets.
In addition, existing learning\text{-}based techniques either only focus on the changed code~\cite{yin2019learning,hoang2020cc2vec,zhou2021finding}, separately encode the changed code and its context~\cite{lin2021context,le2021deep}, or encode the code change as a whole~\cite{jiang2017auto,wang2021context,panthaplackel2021deep}.
Some of them ignore the context or do not highlight the changed code.
All of them lack explicit interaction between the changed code and the whole code change.
These hinder existing techniques from effectively capturing information from code changes.

Only a few studies focus on general approaches for learning code change representations that can be used in diverse tasks~\cite{yin2019learning,hoang2020cc2vec}.
Yin \etal{}~\cite{yin2019learning} proposed to learn distributed representations of small code edits by training an auto-encoder to reconstruct edits.
Unfortunately, their approach only focuses on small code edits (i.e., a single hunk with no more than 3 changed lines) while many software engineering tasks require encoding code changes with multi hunks~\cite{jiang2017auto,zhou2021finding}.
Hoang \etal{}~\cite{hoang2020cc2vec} proposed an approach named CC2Vec to learn code change representations that can be used to boost multiple code-change-related tasks.
However, CC2Vec only considers the added and removed code lines and ignores their context.
Also, it requires commit messages, i.e., natural language descriptions of code changes, to supervise the representation learning process.
However, commit messages are not always available for code-change-related tasks~\cite{tian2020evaluating,lin2021context}, and it is challenging to collect and identify high-quality commit messages~\cite{dyer2013boa,jiang2017auto,liu2018how}.

Considering the lack of general code change representation approaches and to boost existing solutions to code-change-related tasks, this paper proposes a \textbf{C}ode \textbf{C}hange \textbf{Rep}resentation learning approach named \textbf{\toolname{}}, which acts as a general code change encoder and can be used in various downstream tasks.
Compared to Yin et al.'s work~\cite{yin2019learning}, \toolname{} targets commit-level code changes, which are employed by plenty of common code-change-related tasks. 
Different from CC2Vec~\cite{hoang2020cc2vec}, our approach is jointly trained with task-specific components like classifiers on the target task, not requiring additional labels for supervision.
Considering the limitations of existing task-specific learning-based techniques, first, \toolname{} adopts a pre-trained code model to learn the representations of the code before and after a change.
The pre-trained code model can build strong code representations~\cite{codebert2020feng}, forming a good basis for learning code change representations on diverse downstream tasks.
Moreover, a novel mechanism named \textbf{query back} is proposed and used in \toolname{} to highlight the changed code and learn to adaptively select important information from the code change by explicitly interacting the changed code fragments with the whole code change.

Specifically, given a code change, \toolname{} first splits it into the before-change code and the after-change code, compares the two code versions, and records the alignment information between them.
Next, a pre-trained code model is adopted to compute the contextual embeddings of the two code versions, respectively.
Then, the query-back mechanism is leveraged to capture the information related to the changed code from the contextual embeddings.
In detail, it locates the changed code fragments via the alignment information and extracts a feature vector from them to capture change details.
This change-aware feature vector is used as a query to ``retrieve'' related context information from the before-change and after-change code through attention ~\cite{vaswani2017attention}, namely query back, and produce the final code change representation.

To show the effectiveness and the generalization of \toolname{}, we evaluate \toolname{} on three code-change-related tasks: 1) commit message generation (CMG), 2) automated patch correctness assessment (APCA), 3) just-in-time defect prediction (JIT-DP).
Experimental results show that:
on CMG, \toolname{} improves the state-of-the-art approaches by 11.8\% and 12.8\% in terms of BLEU on two datasets.
For APCA, \toolname{} improves the best baseline by 5.0\% and 10.2\% in terms of F1 and AUC, respectively.
On JIT-DP, \toolname{} also outperforms the best-performing baseline by 2.1\%-10.7\% in terms of AUC on five projects.
We also conduct ablation studies, which show that both the pre-trained code model and the query-back mechanism are helpful for each task.

The contributions of this work can be summarized as follows:
\begin{itemize}
    \item We propose the novel query-back mechanism for encoding code changes, which can highlight the changed code and learn to adaptively select important information from a code change.
    \item We propose a novel code change representation approach named \toolname{}, which consists of a pre-trained code model and the query-back mechanism. \toolname{} is plug-and-play and can be used in diverse code-change-related tasks.
    \item We comprehensively evaluate \toolname{} on three downstream tasks. Experimental results show that \toolname{} outperforms the state-of-the-art baselines on each task. 
    \item We provide empirical evidence of the generalizability of pre-trained code models on diverse code-changed-related tasks.
    \item We release our replication package\footnote{\url{https://github.com/ZJU-CTAG/CCRep}}, including our source code and used datasets, for follow-up works.
\end{itemize}

The remainder of this paper is organized as follows:
Section 2 introduces the problem, pre-trained code models and the motivation of the query-back mechanism.
Section 3 elaborates on our approach.
Section 4 presents the procedures and results of our evaluation on three tasks.
We discuss the variants and the limitations of our approach and the threats to validity in Section 5.
After briefly reviewing the related work in Section 6, we conclude and point out future work in Section 7.
 \section{Problem and Preliminary}
This section describes the problem we aim to solve, briefly introduces pre-trained code models and presents the motivation of the query-back mechanism.

\subsection{Problem}
This work focuses on proposing a learning-based code change representation approach, or in other words a code change encoder, that can be used in various code-change-related tasks.
A code change $T$ consists of the code versions before and after the change, i.e., $T^b$ and $T^a$.
Both $T^b$ and $T^a$ consist of a sequence of tokens, i.e., $T^b = [t^b_1, t^b_2, ..., t^b_{|T^b|}]$ and $T^a = [t^a_1, t^a_2, ..., t^a_{|T^a|}]$, where $|x|$ refers to the length of $x$.
A code change encoder can be viewed as a function $f$ which converts $T$ into a numeric vector $v$, i.e., $v = f(T)$.
Because different tasks may value different properties of code changes, in this work, we expect the proposed approach to be jointly trained with task-specific components on the target task.

\ignore{\subsection{From Attention to Transformer}
Attention~\cite{bahdanau2015neural} is widely used in natural language processing (NLP) to help a model capture important information of the inputs~\cite{vaswani2017attention}.
It can be regarded a function that takes as input a query and a set of key-value pairs, which are all vectors, and outputs a weighted sum of the values.
\toolname{} also uses attention to capture important information from code.
Specifically, we use the scaled dot-product attention~\cite{vaswani2017attention}, which is widely-used and performs well.
It works as follows:
\begin{equation}
    \text{Attention}(Q, K, V) = \text{softmax}(\frac{QK^T}{\sqrt{d_k}})V
\end{equation}
where Q is a sequence of queries, K and V denotes keys and values, respectively, and $d_k$ refers to the dimension of each key.

Self-attention is a special case of attention, where $Q=K=V$~\cite{vaswani2017attention}.
It can encode sequence data into contextual embeddings and has become a basic sequence model~\cite{bert2019jacob,codet52021wang}.
To allow models to jointly attend to information from different representation subspaces, researchers proposed multi-head attention, which projects the queries, keys and values multiple times using different, learned linear projections, calculates multiple attentions and concatenates them as one, as follows:
\begin{equation}
    \begin{aligned}
    &\text{MultiHead}(Q, K, V) = [head_1; ...; head_n]W^M,\\
    &head_i = \text{Attention}(QW^Q_i, KW^K_i, VW^V_i)
    \end{aligned}
\end{equation}

Transformer~\cite{vaswani2017attention} is a widely used sequence-to-sequence model, which consists of multiple layers with attention as the core.
Compared to RNN and its variants, Transformer can better handle long sequence and are more parallelizable.
Therefore, it has been used as the backbone network of many large models~\cite{bert2019jacob,t52020reffel,gpt2019radford}.
}

\subsection{Pre-Trained Code Model}
Model pre-training is widely used in the natural language processing (NLP) community and the produced pre-trained models have shown to be effective in various NLP tasks~\cite{bert2019jacob,t52020reffel,gpt2019radford}.
The rationales behind the impressive effectiveness of pre-trained models include: 1) pre-trained models learn high\text{-}quality language representations from huge corpora,
2) pre\text{-}trained models provide good parameter initializations for downstream tasks~\cite{liu2020multi}, and
3) pre-trained models are usually large and can avoid overfitting on the small datasets of downstream tasks.~\cite{liu2020multi}.
Recently, researchers also applied model pre\text{-}training to code and released a number of pre\text{-}trained code models, such as  CodeBERT~\cite{codebert2020feng}, GPT-C~\cite{svyatkovskiy2020intelli}, PLBART~\cite{plbart2021ahmad} and CodeT5~\cite{codet52021wang}.
These models use Transformer-based architectures, can be used to encode and/or generate code or code-related texts, and have been shown to significantly boost code understanding and generation tasks~\cite{codebert2020feng,codet52021wang}.
The impressive performance and generality of these models inspire us to investigate their feasibility in code change representation learning.

\subsection{Motivation of Query-Back Mechanism}

\begin{table}[!t]
\small
    \centering
    \ra{1.3}
    \caption{Motivating Example of Query-Back Mechanism}
    \label{tab:motivating-example}
    \begin{threeparttable}
    \addtolength\tabcolsep{3pt}
\begin{tabular}{|p{0.46\textwidth}@{\hskip3pt}|}
        \hline
        \vspace{-0.4cm}
        \begin{lstlisting}[numbers=left]
@@ -277,8 +277,8 @@ public class EditPost extends Activity {
 } else {

    if (extras != null) {
-       id = WordPress.currentBlog.getId();
        try {
+           id = WordPress.currentBlog.getId();
            blog = new Blog(id, this);
        } catch (Exception e) {
            Toast.makeText(this,
        \end{lstlisting}\vspace{-0.4cm}\\
        \hline
        \textbf{Commit Message}: Moved post id creation to try catch block to help EditPost activity recover if there's no valid currentBlog\\
        \hline
    \end{tabular}
\end{threeparttable}
\end{table}
     A code change contains both the changed code and its context.
Table~\ref{tab:motivating-example} presents a code change with its commit message collected from the WordPress-Android project~\cite{website:wordpress}.
We can see from this example that:
1) The changed code is the core of a code change.
For this example, by inspecting line 5 and line 7, we can know that the developer ``moved post id creation''.
2) The context may provide important information for understanding the code change.
For instance, based on the context in Table~\ref{tab:motivating-example}, we can know that the code change is related to ``try catch block'' and ``EditPost''.
3) Not all the context is useful. For this example, line 2-4, line 8 and line 10 are unrelated to the content and intent of this code change.
As discussed in Section~\ref{sec:intro}, existing code change representation approaches either ignore the context~\cite{yin2019learning,hoang2020cc2vec,zhou2021finding}, do not highlight the changed code~\cite{jiang2017auto,wang2021context,panthaplackel2021deep}, or 
consider all the context without adaptive information selection~\cite{lin2021context,le2021deep}.
These hinder their effectiveness and generality, and motivate us to propose the query-back mechanism to explicitly highlight the changed code and learn to adaptively capture information from the code change.
 \section{Approach}
\begin{figure*}[!t]
    \includegraphics[width=0.92\textwidth]{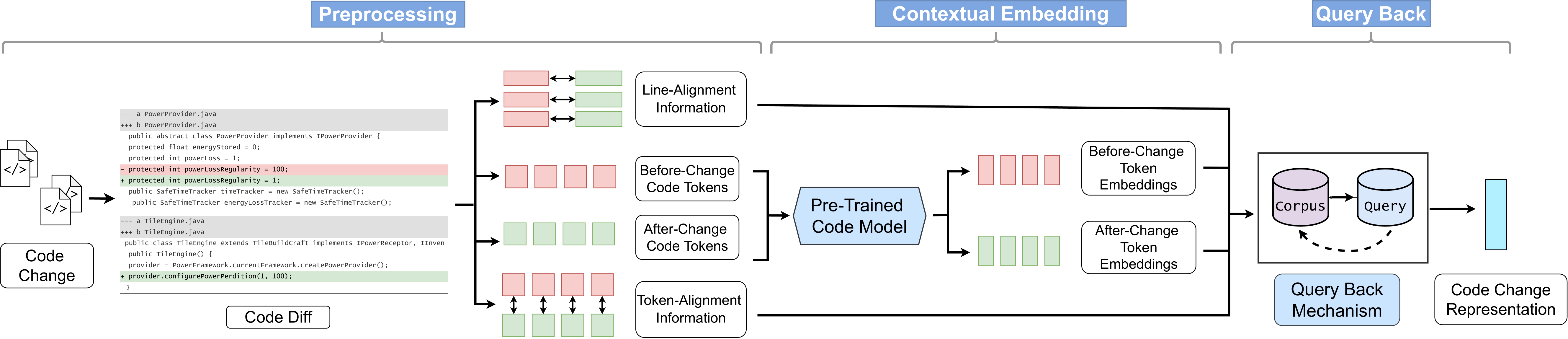}
    \centering
    \caption{The overall framework of \toolname{}.}
    \label{fig:framework}
  \end{figure*} 
The framework of \toolname{} is presented in Figure~\ref{fig:framework}.
\toolname{} takes a code change as input and generates its vector representation.
It consists of three parts:

\noindent\textbf{Code Change Preprocessing}. Given a code change $T$, this part prepares the token sequences of the before-change and after-change code, i.e., $T_b$ and $T_a$, identifies and aligns the modified code fragments in them, and stores their alignment information.

\noindent\textbf{Contextual Code Embedding}. This part leverages a pre-trained code model to obtain the contextual embedding of each token in $T_b$ and $T_a$, respectively.

\noindent\textbf{Query-Back Mechanism}. This part takes as input the contextual embeddings of $T_b$ and $T_a$ and the alignment information of the changed code fragments, leveraging the query-back mechanism to produce the vector representation of $T$.

\subsection{Code Change Preprocessing}\label{sec:prepro}
Given a code change $T$, we first use a text diff tool, such as Python difflib~\cite{website:difflib}, to convert it as a code diff.
A code diff contains one or more hunks and each hunk includes the lines that are deleted from the before-change code (the deleted lines), the lines that are added in the after-change code (the added lines), and the unchanged lines before and after them (the context).
We can extract the changed code fragments at different granularity levels, e.g., changed code tokens or changed code lines, from a code diff.
Then, we split the code diff as a sequence of hunks and preprocess these hunks as follows: 

\begin{table*}[!t]
\centering
\caption{An Illustrative Example of Line Aligning}
\label{tab:line-aligning}
\begin{threeparttable}
\begin{tabular}{|c|l|l|c|c|c|}
\hline 
\textbf{Line Pair Index} & \textbf{Before-Change Code Line \ } & \textbf{After-Change Code Line} & \textbf{Line Change Type} & $l_{i}^{b}$ & $l_{i}^{a}$ \\
\hline 
1 & if (cursor != null) \{ & if (cursor != null \&\& cursor.moveToFirst()) \{ & Replace & 1 & 1 \\
\hline 
2 & cursor.moveToFirst(); & - & Delete & 2 & N/A \\
\hline 
3 & int idx = cursor.getColumnIndex(); & int idx = cursor.getColumnIndex(); & Keep & 0 & 0 \\
\hline 
4 & - & if (idx != -1) & Add & N/A & 4 \\
\hline 
5 & result = cursor.getString(idx);\} & result = cursor.getString(idx);\} & Keep & 0 & 0 \\
\hline
\end{tabular}
\begin{tablenotes}
    \item ``-'' refers to an empty line.
\end{tablenotes}
\end{threeparttable}
\end{table*} 
\textbf{Line Aligning}.
As demonstrated in Table~\ref{tab:line-aligning}, for each hunk, we first align the before-change and after-change code line-by-line using Python difflib and obtain multiple aligned line pairs. A newly added/deleted line is regarded to be aligned with an empty line. We label the lines in each pair with a line index $l_i$ starting from 1. Specifically, for each line pair: 1) If it refers to a line change, i.e., addition, deletion or replacement, we label the lines in the pair with $l_i$. 2) If its two lines are the same, they are both labeled with the default index 0. After processing one aligned line pair, no matter whether this line pair refers to a line change or not, $l_i$ is increased by 1. For example, in Table~\ref{tab:line-aligning}, the first, second, and fourth line pairs refer to line replacement, deletion, and addition, respectively. The lines in them are labeled with 1, 2 and 4, respectively. The lines in the third/fifth line pairs are the same, so they are labeled with 0. When we finish aligning one hunk, the current line index $l_i$ is passed to the next hunk as its initial line index. After this step, every line in both the before-change and after-change code will have a line index, denoted as $l^b_i$ and $l^a_i$.

\textbf{Tokenizing}.
The embedding layer in the pre-trained code model is tightly bound to the vocabulary of the model's tokenizer.
Therefore, to correctly make use of the pre-trained embedding layer, we use the tokenizer provided by the pre-trained code model to tokenize each code line into a token sequence.
Such tokenizer is usually based on subwords, e.g., BPE~\cite{sennrich2016neural}, and needs to build a subword vocabulary from a corpus before pre-training.
It can split a text into subwords and avoid the out-of-vocabulary problem~\cite{karampatsis2020big}.
Besides, for each token, we store the index $l_i$ of the line it belongs to.

\textbf{Flattening}.
To prepare the flat token sequences that can be processed by the pre-trained code model, for each of the before-change and after-change code, we independently collect its tokens from all hunks in the diff and concatenate the tokens into a sequence.
The token sequences of the before-change and after-change code are denoted as $T^b = [t^b_1, t^b_2, ..., t^b_{|T^b|}]$ and $T^a = [t^a_1, t^a_2, ..., t^a_{|T^a|}]$, respectively.

\textbf{Token Aligning}.
We align $T^b$ and $T^a$ token by token using Python difflib to identify the changed tokens. 
After aligning, every token in $T^b$ ($T^a$) will get a token change flag $m^b_i$ ($m^a_i$), which is 1 for changed tokens and 0 for unchanged ones.

\subsection{Contextual Code Embedding}
This part takes as input the token sequences of the before-change and after-change code, i.e., $T^b$ and $T^a$, aiming to independently encode them as two sequences of contextual embeddings, i.e., $H^b = [h^b_1, h^b_2, ..., h^b_{|T^b|}]$ and $H^a = [h^a_1, h^a_2, ..., h^a_{|T^a|}]$.
$H^b$ and $H^a$ are expected to capture the syntactic and semantic information of the code before and after the change.
\toolname{} leverages a pre-trained code model as the code encoder.
Because pre-trained code models are shown to be able to produce high-quality code representations and can be applied to datasets of different sizes~\cite{codebert2020feng,svyatkovskiy2020intelli,codet52021wang}.
Specifically, our implementation of \toolname{} uses CodeBERT, since it is widely used and has been shown to perform well on multiple code-related tasks~\cite{codebert2020feng,zhou2021finding,lin2021traceability}.
Given $T^b$ or $T^a$, CodeBERT uses a multi-layer Transformer~\cite{vaswani2017attention} to make code tokens aggregate context information from each other and outputs their contextual embeddings $H^b$ or $H^a$.
Please note that \toolname{} is agnostic to pre-trained code models and CodeBERT can be substituted with other pre-trained models that can be used as a code encoder.

\subsection{Query-Back Mechanism}\label{sec:query_back}

\begin{figure*}[!t]
    \includegraphics[width=0.95\textwidth]{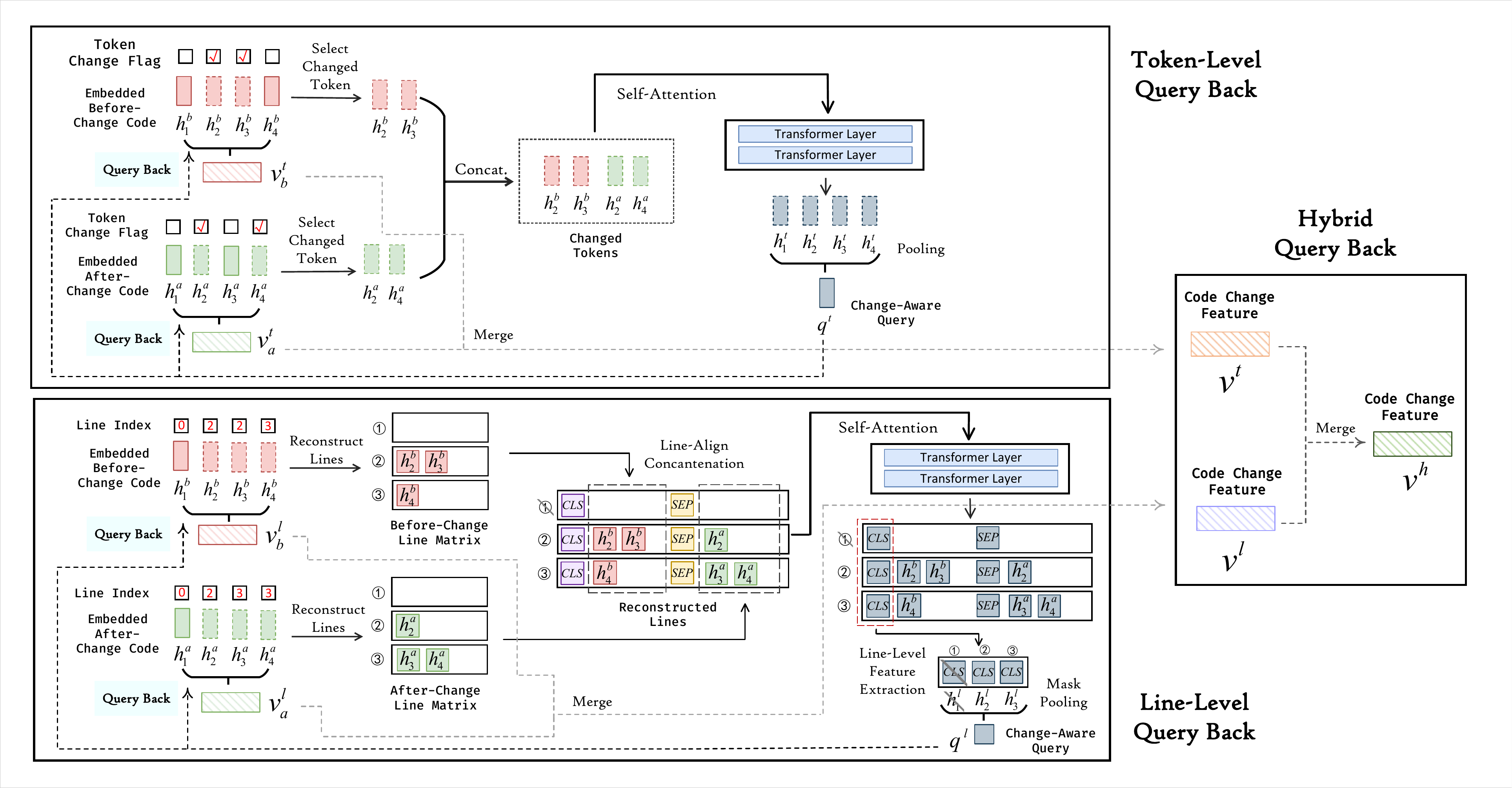}
    \caption{The architectures of the token-level, the line-level, and the hybrid query-back mechanisms. The upper and lower parts depict the token-level and the line-level query-back mechanisms, respectively. They both take as input the contextual embeddings of the before-change and after-change code (the ``corpus''), together with the alignment information, i.e., token change flags and line indices, extracted during preprocessing. Token-level query-back mechanism selects and combines changed tokens based on token change flags, constructs a query $q^t$ using a Transformer and the average pooling, and adaptively retrieves important information from the ``corpus'' to represent the code change as a feature vector $v^t$. Line-level query-back follows a similar procedure, but it reconstructs the structure of changed line pairs based on line indices, explicitly masks unchanged lines when constructing the query $q^l$, and outputs a feature vector $v^l$ as the code change representation. Hybrid query-back combines $v^t$ and $v^l$ to construct the code change representation $v^h$.}
\label{fig:query-back}
\end{figure*} 
This part aims to produce the final representation $v$ of the input code change $T$.
To help \toolname{} effectively capture important information, we propose a novel mechanism named \textbf{query back} for this part.
Its main idea is to encode the changed code fragments as a change-aware query $q$, use such query to ``retrieve'' important information from the before-change and after-change code (the ``corpus''), and produce the final code change representation $v$ based on the ``retrieved'' information.
It can be seen that this ``retrieval'' process makes the changed code explicitly interact with the whole code change, so that we can adaptively extract important information from the ``corpus''.

Specifically, this part takes as input the contextual embeddings $H^b$ and $H^a$ produced by the pre-trained code model and the alignment information extracted during preprocessing, and outputs the code change representation $v$.
Considering that different tasks may focus on the changed code fragments of different granularity, we propose three variants of the query-back mechanism, namely the token-level, the line-level, and the hybrid query-back mechanisms.
Figure~\ref{fig:query-back} shows their architectures.

\subsubsection{Token-Level Query-Back Mechanism}\label{sec:token_level}
This variant constructs the change-aware query $q^t$ based on the changed code tokens, which is beneficial if the changed code is fine-grained, e.g., only changes the name of a method or renames an identifier.
As shown in the upper part of Figure~\ref{fig:query-back}, it works as follows:

\textbf{Changed Token Selection}. During preprocessing, each token in $T_b$ and $T_a$ is assigned a token change flag $m^b_i$ and $m^a_i$.
Based on these flags, we first construct a flag sequence $M^b = [m^b_1,m^b_2,...,m^b_{|T_b|}]$ ($M^a = [m^a_1,m^a_2,...,m^a_{|T_a|}]$) for $T_b$ ($T_a$).
Then, the contextual embeddings of the changed code tokens are picked out from $H^b$ and $H^a$ based on $M^b$ and $M^a$ , and are further concatenated as a new embedding sequence $H'$, as follows:
\begin{equation}
	\begin{aligned}
    [h^{b'}_1,h^{b'}_2,...,h^{b'}_{n_b}] &= [h^b_1,h^b_2,...,h^b_{|T_b|}] \otimes [m^b_1,m^b_2,...,m^b_{|T_b|}],\\
    [h^{a'}_1,h^{a'}_2,...,h^{a'}_{n_a}] &= [h^a_1,h^a_2,...,h^a_{|T_a|}] \otimes [m^a_1,m^a_2,...,m^a_{|T_a|}], \\ 
    H' &= [h^{b'}_1,h^{b'}_2,...,h^{b'}_{n_b},h^{a'}_1,h^{a'}_2,...,h^{a'}_{n_a}],
	\end{aligned}
\end{equation}
where $n_b$ and $n_a$ refer to the numbers of the changed tokens in $T_b$ and $T_a$, $\otimes$ denotes the masked-select operation which selects elements from the left operand based on the flags provided by the right operand.
This step localizes the fine-grained token changes and can help the model concentrate on them during feature extraction.

\textbf{Query Construction}. In this step, we make the changed tokens aware of each other and encode them as a single vector, namely the change-aware query $q^t$.
In detail, we first use a multi-layer Transformer to extract change-aware features $H^t=[h^t_1,h^t_2,...,h^t_{n_b+n_a}]$ from $H'$. Then, average pooling is applied to $H^t$ to squash it into the query $q^t$, as follows:
\begin{equation}
  H^t = \text{Transformer}(H')
  \end{equation}
\begin{equation}
  q^t = \text{Pooling}(h^t_1,h^t_2,...,h^t_{n_b+n_a})
\end{equation}
$q^t$ is expected to encode the information of token changes.

\textbf{Query Back}. The extracted change-aware query $q^t$ is used to ``retrieve'' relevant information from the before-change and after-change code (the ``corpus'') through attention.
Since $q^t$ is also learned from the ``corpus'', we call this attention query-back attention.
Specifically, we adopt the multi-head attention~\cite{vaswani2017attention} to implement the ``retrieval'' process, where both the keys and the values are set to the contextual embeddings $H^b$ or $H^a$ and $q^t$ is used as the query, as follows:
\begin{equation}
v^t_b = \text{MultiHead}(q^t, H^b, H^b)
\end{equation}
\begin{equation}
v^t_a = \text{MultiHead}(q^t, H^a, H^a)
\end{equation}
$v^t_b$ and $v^t_a$ refer to the attended feature vectors retrieved from the before-changed and after-change code, respectively.
\todo{How many heads do we use?}

\textbf{Merging}. Eventually, we merge $v^t_b$ and $v^t_a$ by element-wise addition to get the final code change representation $v^t$:
\begin{equation}
  v^{t} = v^t_b + v^t_a.
\end{equation}

\subsubsection{Line-Level Query-Back Mechanism}
This variant constructs the change-aware query $q^l$ based on the changed code lines, which captures line-level code change features and can be useful if one or more code lines are completely added or deleted.
As shown in the lower part of Figure~\ref{fig:query-back}, it works as follows:

\textbf{Changed Line Selection}.
In this step, based on the line index $l_i$ of each token generated during preprocessing, we first select out the tokens in the changed lines and reconstruct the line structures of the before-change and after-change code $T^b$ and $T^a$, respectively.
The reconstruction process is identical for $T^b$ and $T^a$, and we leverage the example in Figure~\ref{fig:query-back} to illustrate it:
(1) First, we use the line indexes of all the tokens in $T^b$ ($T^a$) to form a line index sequence $[l^b_1,l^b_2,...,l^b_{|T_b|}]$ ($[l^a_1,l^a_2,...,l^a_{|T_a|}]$).
In our example, such sequence is $[0,2,2,3]$ ($[0,2,3,3]$).
(2) Then, we initialize an empty matrix of shape $L\times W \times d$ for reconstructing line structures, where $L$, $W$ and $d$ refer to the maximum lines of the code (line-dimension), the maximum tokens in a code line and the dimension of a contextual embedding. In our example, $L$=3 and $W$=3.
We refer to this matrix as line matrix.
(3) After that, the contextual embedding of each token in $T^b$ ($T^a$) is filled into the corresponding row of the line matrix according to its line index.
In our example, $h^b_2$ and $h^b_3$ are filled into the 2-nd row, while $h^b_4$ is filled into the 3-rd row.
(4) Finally, we order the contextual embeddings in each row by their token indices in $T^b$ ($T^a$). 
In this way, line structures are reconstructed and stored in the line matrix, of which each row stores one line, as shown by the ``Before-Change Line Matrix'' and the ``After-Change Line Matrix'' in Figure~\ref{fig:query-back}.
We refer to the process mentioned above as the scattering-reshaping operation, and briefly formulate it as:
\begin{equation}
  \begin{aligned}
    Ls^b &= [h^b_1,h^b_2,...,h^b_{|T_b|}] \odot [l^b_1,l^b_1,...,l^b_{|T_b|}], \quad Ls^b \in \mathcal{R}^{L \times W \times d}\\
    Ls^a &= [h^a_1,h^a_2,...,h^a_{|T_a|}] \odot [l^a_1,l^a_1,...,l^a_{|T_a|}], \quad Ls^a \in \mathcal{R}^{L \times W \times d}\\ 
\end{aligned}
\end{equation}
where $\odot$ refers to the scattering-reshaping operation, $Ls^b$ and $Ls^a$ refer to the line matrices of $T^b$ and $T^a$, respectively.
Note that the tokens from the unchanged lines (with line index 0) are filled into the 0-th row of the line matrix.
Since we only care the changed lines, we simply drop out the 0-th row.
The line matrices $Ls^b$ and $Ls^a$ are concatenated along the second dimension (the token dimension) to pair aligned lines and form longer lines, and two special tokens \textit{CLS} and \textit{SEP} are respectively inserted into the head and the tail of the before-change line in each pair, as follows:
\begin{equation}
  Ls = [CLS,Ls^b,SEP,Ls^a], \quad Ls \in \mathcal{R}^{L \times (2W+2) \times d}
\end{equation}
We refer to such longer line as paired line.

\textbf{Query Construction}. We also use a multi-layer Transformer to model each paired line and make its tokens interact with each other.
Since the tokens in each paired line are in order, we further adopt the positional encoding~\cite{vaswani2017attention}.
For each paired line, we use the hidden state of its first token, i.e. $CLS$, as its feature vector.
The feature vectors of all paired lines are denoted as $H^l=[h^l_1,h^l_2,...,h^l_L]$. 
Then, we mask the vectors of the unchanged lines and apply average pooling to the masked $H^l$ for obtaining the change-aware query $q^l$.
\begin{equation}
  q^l = \text{MaskPooling}(H^l, mask^l)
\end{equation}
where $mask^l$ indicates the changed lines. $q^l$ is expected to capture the information of the changed lines. 

\textbf{Query Back \& Merging}. These two steps are similar to those of the token-level query-back mechanism. We use $q^l$ as the query to ``retrieve'' information from $H^b$ and $H^a$, and produce the final code change representation $v^l$.

\subsubsection{Hybrid Query-Back Mechanism}
For some tasks concerning both token changes and line additions/deletions, both the token-level and the line-level change information can be beneficial.
Considering this, we further propose the hybrid query-back mechanism, which fuses the token-level and the line-level query-back mechanisms. 
Specifically, we first obtain the code change representations produced by the token-level and the line-level query-back mechanisms, i.e., $v^t$ and $v^l$.
Then, we linearly project them into the same feature space, normalize them with layer normalization~\cite{ba2016layernorm}, and finally merge them through element-wise addition to produce the final code change representation $v^h$:
\begin{equation}
   v^{h} = \text{LayerNorm}(W_t^Tv^{t}) + \text{LayerNorm}(W_l^Tv^{l})
 \end{equation}
where $W_t$ and $W_l$ are learnable parameters of this module.

\subsection{The Usage of \toolname{}}\label{sec:usage}
\begin{figure}[!t]
    \includegraphics[width=0.5\textwidth]{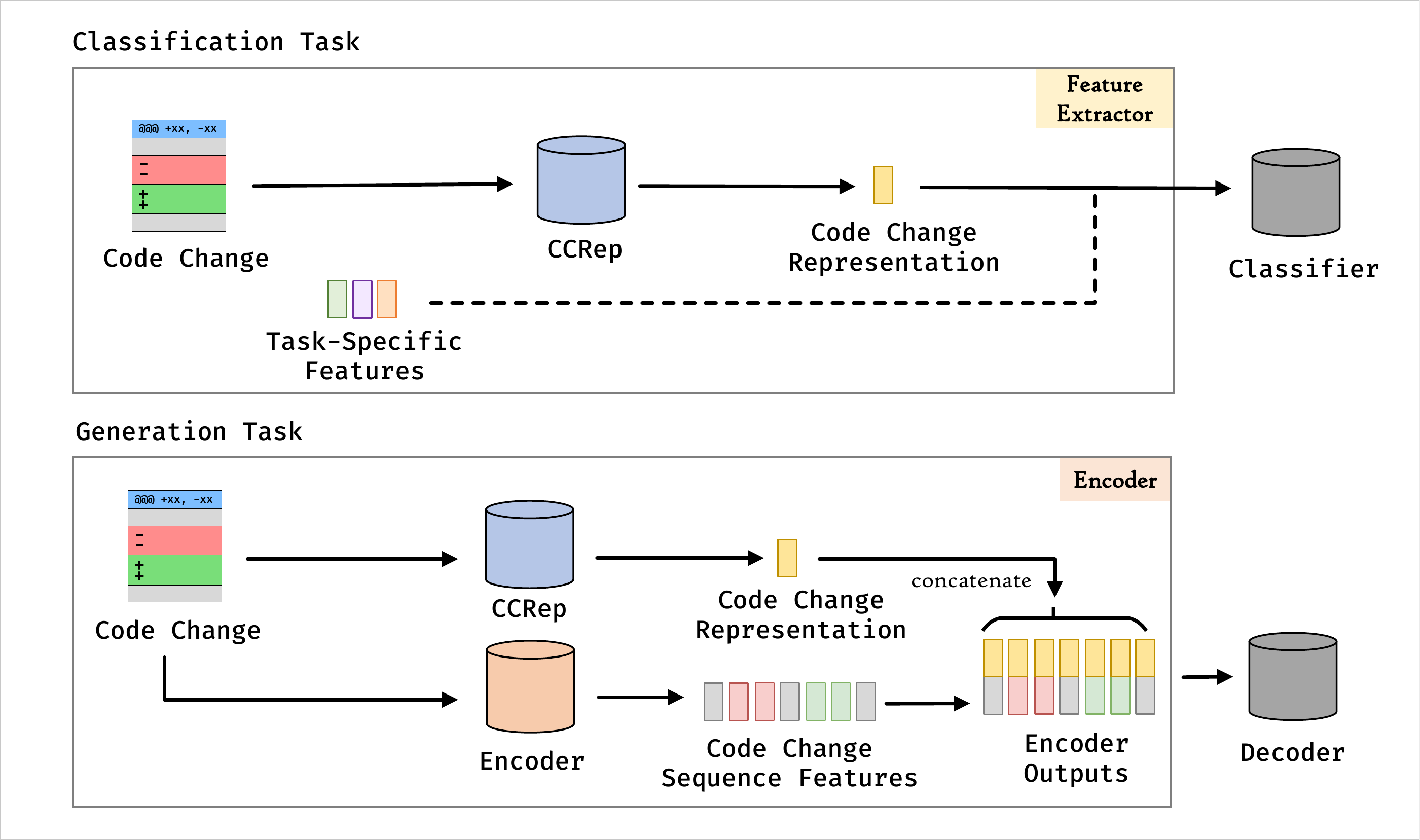}
    \caption{The usage of \toolname{} in code-change-related classification and generation tasks.}
    \label{fig:downstream_usage}
  \end{figure} \toolname{} takes as input a code change and outputs a single numerical vector as the representation of the code change.
It acts as a general encoder and can be used to replace or enhance existing code change encoders in both classification and generation tasks, as shown in Figure~\ref{fig:downstream_usage}.
To use \toolname{} for classification, the most straightforward way is to combine \toolname{} and task-specific components (e.g. a commit message encoder), connect them to a classifier, such as a Multi-Layer Perceptron (MLP), and jointly train \toolname{}, task-specific components and the classifier on the target classification task.
Please note that some tasks only take code changes as input, so there may be no task-specific component.
For generation tasks, \toolname{} can be plugged into a neural generation model to enhance its representations of code changes.
Specifically, given a code change, neural generation models usually leverage a neural component named encoder to encode it into a sequence of feature vectors, and fed such vectors into another neural component named decoder for generation.
We can concatenate the single vector produced by \toolname{} and each feature vector produced by the encoder of a neural generation model, respectively, to enhance the representations of code changes, and feed those enhanced representations into the decoder for better generation.
Also, \toolname{} and the original encoder and decoder are jointly trained on the target generation task, instead of using a two-stage training scheme as CC2Vec.
 \section{Experiments}\label{sec:exp}
To evaluate the effectiveness of \toolname{} and demonstrate its applicability to diverse downstream tasks, we apply \toolname{} to three different tasks related to code changes: commit message generation~\cite{wang2021context}, automated patch correctness assessment~\cite{lin2021context}, and just-in-time defect prediction~\cite{zeng2021deep}.
For each task, we aim to answer two research questions:

\noindent\textbf{RQ1: How effective is \toolname{} compared to the state-of-the-art approaches?}
We evaluate the three variants of \toolname{}, which use the token-level, the line-level and the hybrid query-back mechanisms and are referred to as \varianttoken{}, \variantline{} and \varianthybrid{}, respectively, on the target task, and compare them against the state-of-the-art approaches. 

\noindent\textbf{RQ2: How effective is each component in \toolname{}?}
There are two main components in \toolname{}, i.e., the pre-trained code model and the query-back mechanism.
We conduct ablation studies on each task to investigate their contributions to \toolname{}'s effectiveness.

\subsection{Commit Message Generation}
\subsubsection{Background}
When committing a change into a software repository, developers are encouraged to write a textual message to describe the content and intent of this change, namely, commit message.
Although commit messages are valuable for program comprehension and software maintenance, developers usually neglect writing high-quality commit messages due to time pressure and lack of direct motivation~\cite{dyer2013boa}. 
To alleviate this problem, researchers propose the Commit Message Generation (CMG) task, which takes as input a commit and outputs a concise description summarizing this commit.

\subsubsection{Baselines}
The most recent state-of-the-art CMG approach is FIRA proposed by Dong et.al.~\cite{dong2022fira}, which employs Graph Convolution Network (GCN) \cite{kipf2017semi} as the encoder and a Transformer as the decoder. 
FIRA further proposes a dual copy mechanism to copy both tokens and sub-tokens from the input code change during generation.
According to the evaluation results presented in the FIRA paper, we additionally choose the best-performing retrieval-based CMG approach, i.e., NNGen~\cite{liu2018how}, the best-performing learning-based approach, i.e., \textsc{CoDiSum}~\cite{xu2019commit}, and CoReC~\cite{wang2021context}, which combines retrieval-based and learning-based methods, as the baselines.
We also use LogGen~\cite{hoang2020cc2vec} as a baseline, since it is proposed in the CC2Vec paper and is the adaption of CC2Vec to CMG.

\subsubsection{Our Approach}\label{sec:apca_approach}
Following the description in Section~\ref{sec:usage}, we plug \toolname{} into a neural generation model by concatenating the feature vector produced by \toolname{} with the sequence features of code diff produced by the pre-trained code model (i.e., CodeBERT), as illustrated in the lower part of Figure~\ref{fig:downstream_usage}.

\subsubsection{Experimental Setting}
We evaluate our approach on CMG with the datasets used in the CoReC paper~\cite{wang2021context} and the FIRA paper~\cite{dong2022fira}.
Both datasets are widely-used for this task.
The dataset used by CoReC was initially collected by Jiang et al.~\cite{jiang2017auto} and further cleansed by Liu et al.~\cite{liu2018how}, containing 22.0K commits.
Since \toolname{} focuses on code changes, we filter out the commits with non-code changes, such as binary file changes, file creation or file deletion, resulting in 20.5K commits left.
We re-train and re-evaluate the baselines on our filtered dataset (hereon, the CoReC dataset) using the replication packages provided by their authors~\cite{repo:codisum,repo:nngen, repo:loggen,repo:corec}.
FIRA is not evaluated on this dataset since it needs to parse each code change into two ASTs, but the code diff fragments provided by this dataset are not parsable.
Following CoReC, our approach uses an LSTM as the decoder on this dataset.

The dataset used by FIRA (hereon, the FIRA dataset) was published by Xu et.al.~\cite{xu2019commit}, which contains 75K, 8K and 7.6K commit-message pairs in the training, validation and test sets, respectively. 
Following FIRA, our approach adopts a Transformer as the decoder and also equips with a copy mechanism and the identifier abstraction used by FIRA.
Since Dong et al~\cite{dong2022fira} have evaluated FIRA and other baseline approaches on this dataset, the evaluation results of all the baselines shown in Table~\ref{table-cmg-fira-results} are directly copied from the FIRA paper.

Following prior work~\cite{wang2021context,dong2022fira}, the Adam optimizer~\cite{kingma2015adam} is used to minimize the average cross-entropy loss during training, and BLEU~\cite{papineni2002bleu}, METEOR~\cite{banerjee2005meteor} and ROUGE-L~\cite{lin2004rouge} are used as evaluation metrics.
However, Tao \etal{}~\cite{tao2021on} reported that a variant of the original BLEU called B-Norm BLEU is more correlated with human judgment on the quality of generated commit messages.
Thus, we use the B-Norm BLEU as the substitution of the original BLEU-4 in our experiments (denoted as BLEU as well).

\subsubsection{Results for RQ1}
\begin{table}[]
    \centering
    \caption{CMG: evaluation results on the CoReC dataset}
    \label{table-cmg-results}
\begin{tabular}{@{}cccc@{}}
    \toprule
    \textbf{Model} & \small{BLEU} & \small{METEOR} & \small{ROUGE-L}\\ \midrule
    \footnotesize{LogGen \cite{hoang2020cc2vec}}  & 9.41     & 5.31      & 12.13 \\
    \footnotesize{NNGen \cite{liu2018how}}  & 23.29 & 14.26  & 28.68 \\
    \footnotesize{\textsc{CoDiSum \cite{xu2019commit}}}    & 13.15     & 7.35      & 16.49 \\
    \footnotesize{CoReC \cite{wang2021context}}  & 25.27 & 15.34  & 29.73 \\ \midrule
    \footnotesize{\varianttoken}  & \textbf{28.24} & \textbf{16.99}  & \textbf{34.23} \\
    \footnotesize{\variantline}   & 26.65     & 15.74      & 32.01 \\
    \footnotesize{\varianthybrid} &  27.25    & 16.58      & 33.30 \\ \bottomrule
    \end{tabular}
\end{table} 

\begin{table}[]
    \centering
    \caption{CMG: evaluation results on the FIRA dataset}
    \label{table-cmg-fira-results}
\begin{tabular}{@{}cccc@{}}
    \toprule
    \textbf{Model} & \small{BLEU} & \small{METEOR} & \small{ROUGE-L}\\ \midrule
    \footnotesize{LogGen \cite{hoang2020cc2vec}}  & 8.95     & 8.34      & 10.50 \\
    \footnotesize{NNGen \cite{liu2018how}}  & 9.16     & 9.53      & 11.24 \\
    \footnotesize{\textsc{CoDiSum} \cite{xu2019commit}}    & 16.55     & 12.83      & 19.73 \\
    \footnotesize{CoReC \cite{wang2021context}}  & 13.03 & 12.04 & 15.47 \\
    \footnotesize{FIRA \cite{dong2022fira}}  & 17.67 & 14.93  & 21.58 \\ \midrule
    \footnotesize{\varianttoken}  & \textbf{19.93} & \textbf{16.27}  & \textbf{23.81} \\
    \footnotesize{\variantline}   & 19.79     & 16.06     & 23.60 \\
    \footnotesize{\varianthybrid} &  19.70    & 15.84      & 23.41 \\ \bottomrule
    \end{tabular}
\end{table} Experimental results on CMG are shown in Table \ref{table-cmg-results} and Table \ref{table-cmg-fira-results}.
\varianttoken{}, \variantline{} and \varianthybrid{} stand for the three variants of our approach.
It is shown that our approach outperforms all the baselines in terms of all metrics, indicating the effectiveness of \toolname{} on CMG.
Specifically, on the CoReC dataset, our best-performing variant, i.e., \varianttoken{}, outperforms the best-performing baseline, i.e., CoReC, by 11.8\%, 10.8\% and 15.1\% in terms of BLEU, METEOR and ROUGE-L, respectively.
As for the FIRA dataset, \varianttoken{} outperforms FIRA by 12.8\%, 9.0\% and 10.3\% in terms of the three metrics.
We conduct statistical significance tests using paired bootstrap resampling with 1000 resamples following Koehn~\cite{koehn2004statistical}.
All the p\text{-}values are less than 0.001, indicating that the performance differences between our approach and the baselines are significant.

\subsubsection{Results for RQ2}\label{sec:cmg-ablation-result}
\begin{table}[!t]
  \centering
  \caption{CMG: ablation results on the CoReC dataset}
  \label{table-cmg-ablation}
\begin{tabular}{@{}cccc@{}}
      \toprule
      \textbf{Model} & \small{BLEU}  & \small{METEOR} & \small{ROUGE-L} \\
      \midrule
      \footnotesize{\toolname{} $-$ CodeBERT}  & 27.90 & 16.56  & 33.36 \\
      \footnotesize{\toolname{} $-$ QueryBack}   & 26.23     & 15.72      & 31.78 \\
      \footnotesize{\toolname{}} & \textbf{28.24}     & \textbf{16.99}      & \textbf{34.23} \\ \bottomrule
      \end{tabular}
\end{table} \begin{table}[!t]
  \centering
  \caption{CMG: ablation results on the FIRA dataset}
  \label{table-cmg-fira-ablation}
\begin{tabular}{@{}cccc@{}}
      \toprule
      \textbf{Model} & \small{BLEU}  & \small{METEOR} & \small{ROUGE-L} \\
      \midrule
      \footnotesize{\toolname{} $-$ CodeBERT}  & 18.77 & 15.54  & 22.37 \\
      \footnotesize{\toolname{} $-$ QueryBack}   & 17.88     & 14.78      & 21.33 \\   				
      \footnotesize{\toolname{}} & \textbf{19.93}     & \textbf{16.27}      & \textbf{23.81} \\ \bottomrule
      \end{tabular}
\end{table} To answer RQ2, we compare the best-performing variant of our approach, i.e., \varianttoken{}, with two special models, namely \toolname{}$-$CodeBERT and \toolname{}$-$QueryBack.
The former replaces CodeBERT in \varianttoken{} with a RoBERTa-base model~\cite{liu2019roberta}, which is widely used as a baseline encoder in the NLP community.
\toolname{}$-$QueryBack removes the query-back mechanism from \varianttoken{} and directly feeds the diff's feature vectors produced by the CodeBERT encoder to the decoder for generation.

Experimental results in Table~\ref{table-cmg-ablation} and Table~\ref{table-cmg-fira-ablation} show that our approach outperforms the two special models in terms of all metrics, indicating the effectiveness of the pre-trained code model and the query-back mechanism.
We conduct statistical significance tests like RQ1.
The p-values of \toolname{} compared to \toolname{}$-$QueryBack are all less than 0.001, which means that the query-back mechanism brings statistically significant performance improvements.
The p-values of \toolname{} compared to \toolname{}$-$CodeBERT are all less than 0.05 except the one calculated on the CoReC dataset in terms of BLEU.
This indicates that CodeBERT are also significantly beneficial in most cases.

\subsection{Automated Patch Correctness Assessment}
\subsubsection{Background}
Automated Program Repair (APR)~\cite{goues2019automated} aims to automatically generate bug-fixing patches.
Many APR approaches follow a generate-and-validate methodology, which examines the generated patches with developer-provided test suits.
Because test suits may be inadequate to cover all possible cases, a generated patch that passes all test cases (i.e., plausible patch) may still be incorrect (i.e., overfitting patch).
This is known as the \textit{overfitting} problem of APR~\cite{qi2015analysis,smith2015is}.
To alleviate this problem, many techniques have been proposed to automatically identify correct patches among plausible patches, namely Automated Patch Correctness Assessment (APCA), which is a binary classification task of code changes.

\subsubsection{Baselines}
We consider existing APCA approaches that take patches as input as baselines.
CACHE, recently proposed by Lin \etal{}~\cite{lin2021context}, is the state-of-the-art APCA approach.
Given a patch, CACHE uses AST paths~\cite{alon2019code2vec} to extract the features of its deleted code, its added code and their context, and integrates such features as the patch feature for prediction. 
Another recent work from Tian \etal{}~\cite{tian2020evaluating} leverages pre-trained representation learning models, such as BERT~\cite{bert2019jacob} and CC2Vec~\cite{hoang2020cc2vec}, with some feature comparison functions~\cite{hoang2020cc2vec} to extract patch features and uses classifiers like Logistic Regression (LR) and Decision Tree (DT) for prediction.

\subsubsection{Our Approach}
As illustrated in Section~\ref{sec:usage} and the upper part of Figure~\ref{fig:downstream_usage}, to apply \toolname{} to APCA, we simply connect \toolname{} to a two-layer Multi-Layer Perceptron (MLP) classifier for prediction.

\subsubsection{Experimental Setting}
We use the two datasets constructed by Lin et al.~\cite{lin2021context} for evaluation and refer to them as CACHE-Small and CACHE-Large.
CACHE-Small was constructed by merging and deduplicating the plausible patches collected by Wang \etal{}~\cite{wang2020automated} and Tian \etal{}~\cite{tian2020evaluating}, containing 1,183 patches from the Defects4J benchmark.
CACHE-Large has 49.7K patches and was built by merging and deduplicating the patches from RepairThemAll~\cite{thomas2019empirical} and ManySStuBs4J~\cite{karampatsis2020how}.
Both datasets are roughly balanced.
Five widely-used classification metrics, including Accuracy, Precision, Recall, F1 and AUC, are used for evaluation.
Following Lin \etal{}~\cite{lin2021context}, we perform 5-fold cross-validation on both datasets, use the Adam optimizer to minimize the binary cross-entropy loss during training, and employ Dropout~\cite{srivastava2014dropout} in the classifier.

Because we use the same datasets and experimental settings as Lin \etal{}~\cite{lin2021context}, the evaluation results of the baselines are directly borrowed from their paper.
For space constraints, we only present the variants of Tian \etal{}'s approach that: 1) achieve the best F1 or the best AUC, or 2) use CC2Vec and achieve the best F1 or AUC among all the variants using CC2Vec.

\subsubsection{Results for RQ1}
\begin{table}[!t]
\centering
\caption{APCA: evaluation results on CACHE-Small}
\label{table-apca-cache-small}
\begin{tabular}{@{}cccccc@{}}
\toprule
\textbf{Model}                          & \textbf{Acc.} & \textbf{Pre.} & \textbf{Rec.} & \textbf{F1}   & \textbf{AUC}  \\ \midrule
\footnotesize{LR + CC2Vec \cite{tian2020evaluating}}     & 64.9 & 62.4 & \textbf{90.1} & 73.7 & 68.6 \\
\footnotesize{LR + code2vec \cite{tian2020evaluating}} & 66.8 & 68.6 & 72.9 & 70.6 & 70.2 \\
\footnotesize{CACHE \cite{lin2021context}}                        & 75.4 & \textbf{79.5} & 76.5 & 78.0 & 80.3 \\ \midrule
\footnotesize{\varianttoken}                & 80.5 & 73.8 & 89.0 & 80.6 & 88.1 \\
\footnotesize{\variantline}                 & 81.2 & 75.9 & 86.4 & 80.6 & 88.3 \\
\footnotesize{\varianthybrid}                 & \textbf{82.2} & 76.3 & 88.8 & \textbf{81.9} & \textbf{88.5} \\ \bottomrule
\end{tabular}
\end{table} \begin{table}[!t]
\centering
\caption{APCA: evaluation results on CACHE-Large}
\label{table-apca-cache-large}
\begin{tabular}{@{}cccccc@{}}
\toprule
\textbf{Model}                          & \textbf{Acc.} & \textbf{Pre.} & \textbf{Rec.} & \textbf{F1}   & \textbf{AUC}  \\ \midrule
\footnotesize{DT + BERT \cite{tian2020evaluating}}     & 95.7 & 93.9 & 97.4 & 95.6 & 95.9 \\
\footnotesize{LR + Doc2Vec \cite{tian2020evaluating}} & 90.4 & 91.9 & 88.0 & 89.9 & 96.1 \\
\footnotesize{DT + CC2Vec \cite{tian2020evaluating}} & 95.6 & 95.4 & 95.7 & 95.5 & 95.7 \\
\footnotesize{CACHE \cite{lin2021context}}                                   & 98.6 & 98.9 & 98.2 & 98.6 & 98.9 \\ \midrule
\footnotesize{\varianttoken{}}                                               & 99.4 & 99.6 & 99.2 & 99.4 & 99.97 \\
\footnotesize{\variantline{}}                                                &\textbf{99.7} & 99.8 & \textbf{99.6} & \textbf{99.7} & 99.98 \\
\footnotesize{\varianthybrid{}}                                               & 99.6 & \textbf{99.9} & 99.3 & 99.6 & \textbf{99.99} \\ \bottomrule
\end{tabular}
\end{table}
 
The evaluation results are shown in Table~\ref{table-apca-cache-small} and Table~\ref{table-apca-cache-large}.
It can be seen that on CACHE-Small, our approach outperforms CACHE in terms of all metrics except precision.
Our best variant, i.e., \varianthybrid{}, substantially outperforms CACHE in terms of recall by 12.3 points and improves CACHE in terms of F1 and AUC by 3.9 points and 8.2 points, respectively.
For Tian \etal{}'s approach, our approach outperforms all of its variants, including the CC2Vec-based variants, in terms of F1 and AUC by large margins.

On CACHE-Large, our approach also outperforms Tian et al.'s approach and CACHE in terms of all metrics.
Our best variant, i.e., \varianthybrid{}, achieves an F1 of 99.6\% and an AUC of 99.99\%.
One possible reason behind the impressive performance of both CACHE and our approach is that CACHE-Large is way larger than CACHE-Small, which may ease the learning.
Another possible reason is that CACHE-Large is synthetic to some extent.
Specifically, according to Lin et al.~\cite{lin2021context}, in CACHE-Large, almost all correct patches are human-written patches, while all overfitting patches are generated by APR tools.
Therefore, an approach can perform well if it is able to tell from human-written patches and APR-generated patches.

Another thing worth mentioning is that \varianthybrid{} outperforms the other two variants on the two datasets in terms of AUC, which highlights that both the token-level and the line-level change information is useful for APCA.

\subsubsection{Results for RQ2}
\begin{table}[!t]
\centering
\caption{APCA: evaluation results of the ablation studies}
\label{table-apca-ablation}
\resizebox{\columnwidth}{!}{
\begin{tabular}{@{}ccccccc@{}}
\toprule
\textbf{Dataset}                                                                 & \textbf{Model}                 & \textbf{Acc.} & \textbf{Pre.} & \textbf{Rec.} & \textbf{F1}   & \textbf{AUC}  \\ \midrule
\multirow{3}{*}{\begin{tabular}[c]{@{}c@{}}CACHE-\\ Small\end{tabular}} & \footnotesize{\toolname{} $-$ CodeBERT} & 77.5 & 72.6 & 81.6 & 76.6 & 86.1 \\
                                                                        & \footnotesize{\toolname{} $-$ QueryBack}     & 79.3 & 74.4 & 83.2 & 78.4 & 87.0 \\
                                                                        & \footnotesize{\toolname{}}        & \textbf{82.2} & \textbf{76.3} & \textbf{88.8} & \textbf{81.9} & \textbf{88.5} \\ \midrule
\multirow{3}{*}{\begin{tabular}[c]{@{}c@{}}CACHE-\\ Large\end{tabular}} & \footnotesize{\toolname{} $-$ CodeBERT} & 99.6 & 99.7 & 99.6 & 99.6 & \textbf{99.99} \\
                                                                        & \footnotesize{\toolname{} $-$ QueryBack}     & \textbf{99.7} & 99.8 & \textbf{99.7} & \textbf{99.7} & \textbf{99.99} \\
                                                                        & \footnotesize{\toolname{}}        & 99.6 & \textbf{99.9} & 99.3 & 99.6 & \textbf{99.99} \\ \bottomrule
\end{tabular}
}
\end{table} We compare the best-performing variant, i.e., \varianthybrid{}, with \toolname{}$-$CodeBERT and \toolname{}$-$QueryBack. 
\toolname{}$-$CodeBERT replaces CodeBERT in \varianthybrid{} with the RoBERTa-base model~\cite{liu2019roberta}, which is a widely-used baseline encoder.
\toolname{}$-$QueryBack directly uses CodeBERT to encode a patch's diff and uses the contextual embedding of a special token \textit{CLS} inserted at the beginning of the diff as the code change representation.
Table~\ref{table-apca-ablation} presents the results, which indicate that both the pre-trained code model and the query back mechanism positively affect the effectiveness of \toolname{} on CACHE-Small.
However, there are no significant performance differences between the two special models and \toolname{} on CACHE-Large.
A possible explanation for this is that distinguishing correct and overfitting patches in CACHE-Large is not very hard and less powerful models can also fit the data.

\subsection{Just-in-Time Defect Prediction}
\subsubsection{Background}
Software defects are inevitable and may substantially affect businesses and even people's lives~\cite{hoang2019deepjit}.
On the other hand, the size and complexity of modern software systems grow significantly, making it hard and costly to find defects from them.
To this end, just-in-time defect prediction (JIT-DP) has been proposed to identify defective code changes and provide in-time feedback when developers commit changes to the code base~\cite{kim2008classifying,kamei2016studying,hoang2019deepjit,zeng2021deep}.
Given a commit, JIT-DP targets at predicting whether it is defective and is a binary classification task just like APCA.
However, different from APCA, JIT-DP targets real-world defective commits and can take as input both the code change and the commit message in a commit.

\subsubsection{Baselines}
We use three state-of-the-art JIT-DP approaches, i.e., DeepJIT~\cite{hoang2019deepjit}, CC2Vec~\cite{hoang2020cc2vec} and LAPredict~\cite{zeng2021deep} as baselines.
Given a commit, DeepJIT leverages CNNs to extract feature vectors from the code change and the commit message, respectively, and concatenates such vectors as the commit vector for prediction.
Hoang \etal{}\cite{hoang2020cc2vec} use CC2Vec to encode a code change into a feature vector and appends such vector to the commit feature extracted by DeepJIT for prediction.
We also refer to this approach as CC2Vec for convenience.
LAPredict, proposed by Zeng \etal{}~\cite{zeng2021deep}, uses the number of added code lines as the commit feature and adopts an LR classifier to perform JIT-DP.

\subsubsection{Our Approach}
We follow Section ~\ref{sec:usage} and the upper part of Figure ~\ref{fig:downstream_usage} to apply CCRep in JIT-DP, similar to APCA. 
The only difference is that we concatenate the output of \toolname{} with the commit message feature produced by a CNN (i.e., the task-specific features in Figure~\ref{fig:downstream_usage}) for prediction, following DeepJIT~\cite{hoang2019deepjit}.

\subsubsection{Experimental Setting}
We compare \toolname{} with the baselines on the dataset constructed by Zeng \etal{}~\cite{zeng2021deep}.
The dataset contains six large-scale projects, i.e., OpenStack, QT, Go, Gerrit, Platform and JDT, covering different programming languages.
Considering that CodeBERT is not pre-trained on C++ code~\cite{model:codebert}, we exclude QT for evaluation.
We conduct our experiments following the within-project setting used by Zeng \etal{}~\cite{zeng2021deep} and directly borrow the evaluation results of the baselines from the LAPredict paper.
Since the dataset is highly unbalanced, we follow prior work~\cite{zeng2021deep,hoang2020cc2vec,hoang2019deepjit} and also use AUC as the evaluation metric.
Also following prior work~\cite{hoang2019deepjit}, we use the Adam optimizer to minimize the binary cross-entropy loss and employ Dropout in the classifier.

\subsubsection{Results for RQ1}
\begin{table}[]
\centering
\caption{JIT-DP: evaluation results in terms of AUC}
\label{table-jit-result}
\resizebox{\columnwidth}{!}{
\begin{tabular}{@{}ccccccc@{}}
\toprule
\textbf{Model} & \footnotesize{\textbf{OpenStack}} & \footnotesize{\textbf{JDT}} & \footnotesize{\textbf{Go}} & \footnotesize{\textbf{Platform}} & \footnotesize{\textbf{Gerrit}} & \footnotesize{\textbf{Mean}} \\ \midrule
DeepJIT \cite{hoang2019deepjit}   & 71.32 & 67.01 & 68.91 & 77.12  & 70.25 & 70.92 \\
CC2Vec \cite{hoang2020cc2vec}    & 72.27 & 66.53 & 69.17 & 76.13 & 69.86 & 70.79 \\
LAPredict \cite{zeng2021deep} & 74.91 & 67.57 & 68.31 & 74.61 & 74.95 & 72.07 \\ \midrule
\varianttoken{}   &  75.37    & 68.11     & \textbf{75.63}    &  81.91     & 76.89  & 75.58  \\
\variantline{}    &   \textbf{76.45}   & \textbf{68.96}     & 75.60     & 82.08     & \textbf{77.35}  & \textbf{76.09}  \\
\varianthybrid{}    &  75.63    & 66.63     & 75.48     & \textbf{82.18}     &  77.01 & 75.39   \\ \bottomrule
\end{tabular}
}
\end{table} Experimental results on JIT-DP are summarized in Table~\ref{table-jit-result}.
We can see that \variantline{} is the best-performing variant on average, but the performance differences among the three variants are small.
\variantline{} improves LAPredict, CC2Vec and DeepJIT by 5.6\%, 7.4\% and 7.3\% on average, indicating the effectiveness of \toolname{} in representing code changes for JIT-DP.
We also notice that different projects prefer different query-back mechanisms, which may indicate the different defect characteristics of different projects.

\subsubsection{Results for RQ2}\label{sec:jit-ablation}
\begin{table}[]
\centering
\caption{JIT-DP: ablation results in terms of AUC}
\label{table-jit-ablation}
\resizebox{\columnwidth}{!}{
\addtolength\tabcolsep{0pt}
\begin{tabular}{@{}ccccccc@{}}
\toprule
\textbf{Model} & \footnotesize{\textbf{OpenStack}} & \footnotesize{\textbf{JDT}} & \footnotesize{\textbf{Go}} & \footnotesize{\textbf{Platform}} & \footnotesize{\textbf{Gerrit}} & \footnotesize{\textbf{Mean}} \\ \midrule
\footnotesize{\toolname{} $-$ CodeBERT} & 75.78 & 68.21 & 75.04 & 81.15 & 74.89 & 75.01 \\
\footnotesize{\toolname{} $-$ QueryBack} & 75.70 & 66.69 & \textbf{76.30} & \textbf{82.28} & 74.82 & 75.16 \\
\footnotesize{\toolname{}} & \textbf{76.45} & \textbf{68.96} & 75.63 & 82.18 & \textbf{77.35} & \textbf{76.09} \\ \bottomrule
\end{tabular}
}
\end{table}
 We also conduct ablation studies on JIT-DP to answer RQ2. 
Similar to what we do in APCA, we build and evaluate two special approaches, i.e., \toolname{}$-$CodeBERT and \toolname{}$-$QueryBack. 
As shown in Table~\ref{table-jit-ablation}, the average performance of our approach degrades without CodeBERT or QueryBack, highlighting the effectiveness of the two components in \toolname{}.
We also notice that on Go and Platform, our approach is slightly worse than \toolname{}$-$QueryBack.
After manual inspection, we speculate that this is because a significant number of commits in Go and Platform contain almost only deleted or added code with little context, where the query-back mechanism may bring no benefit.
However, since the performance of our approach and \toolname{}$-$QueryBack is very close, we argue the query-back mechanism can be viewed as harmless on Go and Platform.
In summary, both the pre-trained code model and the query-back mechanism contribute to the effectiveness of \toolname{}.
 
\section{Discussion}
\ignore{\subsection{Comparison with CC2Vec}
It is worth mentioning that we have already explicitly compare with CC2Vec on each task.
Specifically, LogGen is proposed by the paper of CC2Vec~\cite{hoang2020cc2vec} for CMG.
In APCA and JIT-DP, we compare \toolname{} with LR+CC2Vec and CC2Vec, respectively.
On all three tasks, \toolname{} outperforms CC2Vec by substantial margins.}

\subsection{The Variants of \toolname{}}
We propose three variants of \toolname{}, i.e., \varianttoken{}, \variantline{} and \varianthybrid{}.
We can see from Section~\ref{sec:exp} that different tasks prefer different variants.
On CMG, \varianttoken{} performs best, which is probably because that a commit message is generated token by token and the key words in it can often be found from the changed code tokens.
For APCA, the best-performing variant is \varianthybrid.
After inspecting the dataset and the results, we think this is reasonable because a generated patch can either change a few tokens (e.g., replace ``=='' with ``!=''), or insert/delete several lines (e.g., insert a \texttt{NullPointer} check).
As for JIT-DP, \variantline{} is the best.
Based on our inspection, a possible reason is that a defective commit often adds, deletes or modifies multiple lines instead of only a few tokens.
Based on these findings, before applying \toolname{} to other tasks, we encourage the user to analyze the characteristics of the target task and her dataset first and choose the most suitable variant.
On the other hand, on each task, the three variants all outperform the state-of-the-art baselines.
This indicates our approach's effectiveness and makes us believe that any of the variants can serve as a strong baseline for code-change-related tasks.

\subsection{Limitations}\label{sec:limitations}
As discussed in Section~\ref{sec:jit-ablation}, if a code change contains little context code, the query-back mechanism may bring no benefit.
Because in such situation, the changed code and the whole code change contain the same information.
Another limitation of \toolname{} is that the lengths of the code changes that can be processed by \toolname{} are limited by the pre-trained code model.
In detail, the length of the before-change or the after-change code should not exceed the length limit of the used pre-trained model, which is usually 512. 
Fortunately, committing small and coherent code changes has become a widely-recognized good practice, and many techniques have been proposed to decompose tangled code changes~\cite{barnett2015helping,dias2015untangling,tao2015partitioning,wang2019cora,shen2021smart}.
Such practice and techniques can help reduce long code changes and alleviate this limitation.

\subsection{Threats to Validity}
Threats to internal validity refer to the errors and bias in our experiments.
To mitigate such threats, for each task, we try our best to follow the settings used by the state-of-the-art baselines, re-use the evaluation results reported by prior work when possible and use the existing implementations of the baselines to conduct experiments.
We have double checked our code and data and made them publicly available.
Threats to external validity concern the generalization of \toolname{}.
Although we have applied \toolname{} to three different tasks and achieved superior performance, we cannot claim that \toolname{} can be applied to or perform well on all code-change-related tasks.
However, the three tasks have either different inputs or different outputs, care about diverse characteristics of code changes, and cover synthetic and manually-written patches, various projects and multiple programming languages.
Therefore, we believe this threat is limited.
In addition, based on our evaluation results, we argue that \toolname{} can at least serve as a strong baseline to help better solve code-change-related problems.
To minimize threats to construct validity, we choose evaluation metrics following previous studies.
 \section{Related Work}
\ignore{\subsection{Pre-training for Code}
Inspired by the impressive success of pre-training in the natural language processing domain, such as BERT~\cite{bert2019jacob}, GPT~\cite{gpt2019radford} and T5~\cite{t52020reffel}, pre-training has been explored by some recent studies to boost source-code-related tasks~\cite{kanade2020learning,codebert2020feng,graphcodebert2021guo,plbart2021ahmad,codet52021wang}.
These studies pre-trained large models on massive unlabeled code data and fine-tuned the pre-trained models on downstream tasks for better performance.
Feng \etal{}~\cite{codebert2020feng} presented a Transformer-based bimodal pre-trained model called CodeBERT for learning general-purpose representations of programming language (PL) and natural language (NL). 
They used two tasks named Masked Language Modeling (MLM) and Replaced Token Detection (RTD) as their training objectives during pre-training.
Following this work, Guo \etal{}~\cite{graphcodebert2021guo} presented GraphCodeBERT.
GraphCodeBERT uses data flow graph of code snippets, which encodes semantic relation between variables, as extra input during pre-training.
They also developed Edge Prediction and Node Alignment as new pre-training tasks for GraphCodeBERT to exploit information in data flow graph.
Svyatkovskiy \etal{}~\cite{svyatkovskiy2020intelli} pre-trained a variant of GPT-2~\cite{radford2019language} on a large multilingual code dataset, namely GPT-C, and proposed and deployed a code sequence completion system named IntelliCode Compose based on GPT-C.
Liu \etal{}~\cite{liu2020multi} also proposed a Transformer-based pre-trained model named CugLM, to boost code completion.
The pre-training tasks of CugLM include Masked Identifier Prediction, Next Code Segment Prediction and Unidirectional Language Modeling.
In addition, Ahmad \etal{}~\cite{plbart2021ahmad} emphasized the necessity of a pre-training decoder and further presented a unified PL-NL pre-trained model called PLBART.
They pre-trained a denoising auto-encoder by recovering original input from noisy input as BART~\cite{bart2020lewis} on an extensive collection of Java and Python functions and natural language descriptions from GitHub and StackOverflow.
Wang \etal{}~\cite{codet52021wang} proposed a pre-trained model for code understanding and generation built on the T5~\cite{t52020reffel} architecture, namely CodeT5.
CodeT5 extended the idea of sequence-to-sequence pre-training by using novel pre-training tasks such as Masked Span Prediction (MSP) to enable model become aware of identifiers in code.

These studies focus on pre-training language models or sequence-to-sequence models for source code.
In contrast, we aims to learn better code change representations based on existing pre-trained code models for downstream tasks.
Thus, the work on code pre-training is complementary to this work.

\subsection{Learning Code Change Representations}}
The majority of the studies related to code change representation target a specific downstream task and learn code change representations through a task-specific architecture.
Some of them flatten code or code changes as token sequences for representation learning~\cite{jiang2017auto,xu2019commit,wang2021context,liu2020automating,panthaplackel2020learning,hoang2021patchnet,zhou2022spi}.
For example, for commit message generation, applying RNN like LSTM~\cite{hochreiter1998lstm} on diffs is a widely-used approach to extract code change features~\cite{jiang2017auto,xu2019commit,wang2021context}.
To automate comment updates with code changes, Liu \etal{}~\cite{liu2020automating} aligned the tokens of the before-change and after-change code to form an edit sequence and fed such sequence into an LSTM-based encoder to obtain code change representations.
To identify security patches, Zhou \etal{}~\cite{zhou2022spi} utilized two LSTMs to respectively learn the statement-level features of the added and deleted code in a patch and merged their features with a multi-layer convolutional neural network.

Some studies leverage the syntactic structure of code, e.g., AST, to enhance code change representation learning~\cite{liu2020atom, brody2020structural, lin2021context, yin2019learning, panthaplackel2021deep}.
For example, to assess patch correctness, Lin \etal{}~\cite{lin2021context} proposed to extract and encode the changed AST paths and the unchanged AST paths, respectively, and merge their feature vectors as the code change representation.
Yin \etal{}~\cite{yin2019learning}, Panthaplackel \etal{}~\cite{panthaplackel2021deep} and Dong \etal{}~\cite{dong2022fira} proposed to converted the two ASTs of a code change into a graph and use graph neural networks, e.g., GGNN~\cite{li2016gated} and GCN~\cite{kipf2017semi}, to learn code change representations from the graph.

Recently, several studies adopted pre-trained code models to represent code changes for specific downstream tasks~\cite{zhou2021finding,tian2020evaluating,lin2021traceability,zhou2021assessing}.
For instance, to identify silent vulnerability fixes, 
Zhou \etal{}~\cite{zhou2021finding} leveraged CodeBERT~\cite{codebert2020feng} to encode each changed file in a commit and merged the feature vectors of all changed files as the code change representation.
Lin \etal{}~\cite{lin2021traceability} leveraged pre-trained code models to encode commits for recovering links between issues and commits.
Zhou \etal{}~\cite{zhou2021assessing} also investigated the generalizability of CodeBERT on JIT-DP.
However, they only considered one classification task and only used the data from two projects for evaluation.

Only a few studies aim at learning general-purpose code change representations~\cite{yin2019learning, hoang2020cc2vec}.
Yin \etal{}~\cite{yin2019learning} proposed to learn distributed representations of code edits by training an auto-encoder to reconstruct code edits.
Hoang \etal{}~\cite{hoang2020cc2vec} proposed CC2Vec, which leverages a hierarchical attention network and multiple comparison functions to learn code change representations.
As discussed in Section~\ref{sec:intro}, Yin \etal{}'s approach only focuses on small code edits (i.e., a single hunk with no more than 3 changed lines), while \toolname{} targets commit-level changes.
Also, it lacks explicit interaction between the changed code and the whole code change.
We have tried to compare the performance of Yin \etal{}'s approach with that of \toolname{}.
However, neither Yin \etal{} evaluated their approach on the three tasks used in this work, nor they made their implementation publicly available.
CC2Vec only considers the changed code and ignores the context, and it requires commit messages as labels, which are not always available.
Besides, our evaluation results show that CCRep outperforms CC2Vec on the three tasks by substantial margins.

In summary, our work differs from prior work in several folds:
First, our approach acts as a general code change encoder and can be used in diverse code-change-related tasks.
Second, our approach is equipped with a pre-trained code model and the query-back mechanism, and is technically different.
Third, our evaluation results show that our approach outperforms the state-of-the-art techniques on three tasks.
In addition, this work also investigates the generalizability of pre-trained code models on diverse code-change-related tasks.
 \section{Conclusion}
We propose a novel approach named \toolname{} to learn code change representations.
It acts as a code change encoder and can be jointly trained with and used in diverse code-change-related tasks.
\toolname{} leverages a pre-trained code model to obtain high-quality contextual embeddings and better handle datasets of different sizes, and a novel mechanism named query back to highlight the changed code and adaptively capture related context information.
We evaluate \toolname{} on one generation task and two classification tasks.
Experimental results show that \toolname{} outperforms the state-of-the-art approaches on each task and both the pre-trained code model and the query-back mechanism contribute to its effectiveness.
In the future, we plan to apply our approach to more code-change-related tasks and improve it to encode long and structured code changes.

\section{Acknowledgments}
This research/project is supported by the National Natural Science Foundation of China (No. 62202420) and the Fundamental Research Funds for the Central Universities (No. 226-2022-00064). Zhongxin Liu gratefully acknowledges the support of Zhejiang University Education Foundation Qizhen Scholar Foundation.

\balance
\bibliographystyle{IEEEtran}
\bibliography{references}

\end{document}